\documentstyle[twoside,fleqn]{article}

\def\noi{\noindent}

\makeatletter
\renewcommand{\section}{\@startsection{section}{1}{0pt}%
        {-3.5ex plus -1ex minus -.2ex}{2.3ex plus .2ex}%
        {\large\bf\protect\raggedright}}

\renewcommand{\subsection}{\@startsection{subsection}{2}{0pt}%
        {-3ex plus -1ex minus -.2ex}{1.4ex plus .2ex}%
        {\normalsize\bf\protect\raggedright}}

\renewcommand{\thesubsubsection}%
        {\arabic{section}.\arabic{subsection}.\arabic{subsubsection}.}
\renewcommand{\@oddhead}{\raisebox{0pt}[\headheight][0pt]{%
   \vbox{\hbox to\textwidth{\rightmark \hfil \rm \thepage \strut}\hrule}}}
\renewcommand{\@evenhead}{\raisebox{0pt}[\headheight][0pt]{%
   \vbox{\hbox to\textwidth{\thepage \hfil \leftmark \strut}\hrule}}}
\newcommand{\heads}[2]{\markboth{\protect\small\it #1}{\protect\small\it #2}}

\makeatother

\newcommand{\Title}[1]{\noi {\Large #1} \\}
\newcommand{\Author}[2]{\noi{\large\bf #1}\\[2ex]\noi{\it #2}\\}
\newcommand{\Abstract}[1]{\vskip 2mm \begin{center}
     \parbox{16.4cm}{\small\noi #1} \end{center}\bigskip}
\newcommand{\foom}[1]{\protect\footnotemark[#1]}
\newcommand{\foox}[2]{\footnotetext[#1]{#2}}
\newcommand{\email}[2]{\footnotetext[#1]{e-mail: #2}}

\newcommand{\sect}[1]{Sec.\,#1}

\def\nq{\hspace{-1em}}
\def\nqq{\hspace{-2em}}
\def\nhq{\hspace{-0.5em}}
\def\nhh{\hspace{-0.3em}}
\def\cm{\hspace{1cm}}

\newcommand{\Eq}[1]{Eq.\,(\ref{#1})}

\def\eqs{Eqs.\,}
\def\beq{\begin{equation}}
\def\eeq{\end{equation}}
\def\bear{\begin{eqnarray}}
\def\al{&\nhq}
\def\lal{&&\nqq {}}               
\def\bearr{\begin{eqnarray} \lal}
\def\ear{\end{eqnarray}}
\def\earn{\nonumber \end{eqnarray}}
\def\dst{\displaystyle}
\def\tst{\textstyle}

\def\nn{\nonumber\\ {}}
\def\nnv{\nonumber\\[5pt] {}}
\def\nnn{\nonumber\\ \lal }

\def\yy{\\[5pt]}
\def\yyy{\\[5pt] \lal}
\def\eql{\al =\al}

\def\e{{\,\rm e}}
\def\d{\partial}

\def\sign{{\,\rm sign\,}}
\def\diag{{\,\rm diag\,}}
\def\dim{{\,\rm dim\,}}
\def\const{{\rm const}}
\def\Half{{\dst\frac{1}{2}}}
\def\half{{\tst\frac{1}{2}}}

\newcommand{\vars}[1]{\left\{\begin{array}{ll}#1\end{array}\right.}
\newcommand{\lims}[1]{\mathop{#1}\limits}

\def\wider{\vphantom{\int}}

\def\m{{\rm m}}
\def\s{{\rm s}}
\def\qo{\overline{q}}
\def\umx{u_{\max}}
\def\sumi{\sum_{i=1}^{s}}

\def\ds{ds^2_D}
\def\phim{$\phi^{\min}$}
\def\F{\raisebox{0.2ex}{$\lims F_{\rm M}$}{}}
\newcommand{\T}[1]{\raisebox{0.2ex}{$\lims{T}_{#1}$}{}}
\def\mO{\mbox{[-1]}}
\def\pO{\mbox{[1]}}
\def\ss{\scriptscriptstyle}

\def\sph{spherically symmetric\ }
\def\bh{black hole}

\def\RN{Reissner--Nordstr\"om\ }

\heads{Kirill A. Bronnikov and J\'ulio C\'esar Fabris}
{Multidimensional Reissner--Nordstr\"om Problem
		 with a Generalized Maxwell Field}

\begin{document}
\thispagestyle{empty}
\twocolumn[
\noi \unitlength=1mm
\begin{picture}(174,8)
       \put(31,8){\shortstack[c]
       {RUSSIAN GRAVITATIONAL SOCIETY\\
       INSTITUTE OF METROLOGICAL SERVICE \\
       CENTER OF GRAVITATION AND FUNDAMENTAL METROLOGY }     }
\end{picture}
\begin{flushright}
                                         RGS-VNIIMS-96/07      \\
                                         gr-qc/9703022         \\
			       {\it Grav. and Cosmol.} {\bf 2}, 4, 306--314 (1996)
\end{flushright}

\Title{MULTIDIMENSIONAL REISSNER--NORDSTR\"OM PROBLEM \yy
	  WITH A GENERALIZED MAXWELL FIELD\foom 1}

\Author{Kirill A. Bronnikov\foom 2 and J\'ulio C\'esar Fabris\foom 3}
{Departamento de F\'{\i}sica, Universidade Federal do Esp\'{\i}rito Santo,
 Vit\'oria CEP29060-900, Esp\'{\i}rito Santo, Brasil}

{\it Received 22 October 1996}

\Abstract
{We obtain and study static, \sph solutions for
the Einstein --- generalized Maxwell field system in $2n$ dimensions,
with possible inclusion of a massless scalar field.  The generalization
preserves the conformal invariance of the Maxwell field in higher dimensions.
Almost all solutions exhibit naked singularities, but there are some classes
of \bh\ solutions.  For these cases the Hawking temperature is found and its
charge/mass and dimension dependence is discussed. It is shown that, unlike
the previously known multidimensional \bh\ solutions, in our case the \bh\
temperature may infinitely grow in the extreme case (that of minimum mass
for given charges).}

] 
\foox 1
{Talk presented at the 9th Russian Gravitational Conference,
     Novgorod, 24--30 June 1996.}
\foox 2
{Permanent address: Center for Gravitation and Fundamental Metrology, VNIIMS,
     3--1 M. Ulyanovoy Str., Moscow 117313, Russia;
	e-mail: \ kb@goga.mainet.msk.su}
\email 3 {fabris@cce.ufes.br}

\section{Introduction}

Coupling between gravity and gauge fields is a common feature of
many unification theories in higher dimensions \cite{1}.
In supergravity theories, such gauge fields are frequently essential in
order to complete the multiplet structure and to guarantee the invariance of
the Lagrangian with respect to local supersymmetry transformations \cite{2}.
They can provide, in some cases, a dynamical mechanism for the
compactification of extra dimensions \cite{3}. In this supersymmetric
context, they can lead to a geometric interpretation in the superspace.

The simplest example of coupling of a gauge field to gravity is the
Einstein-Maxwell system in four dimensions, whose extension to higher
dimensions can lead to some interesting features \cite{3}: its reduction to
four dimensions implies a non-trivial coupling of the electromagnetic and
scalar fields to gravity.

In the conventional, straightforward generalization of the Maxwell
field to higher dimensions one essential property of the electromagnetic
field is lost, namely, conformal invariance.  The latter exists only in four
dimensions.  There is, however, a generalization which
preserves conformal invariance \cite{4}; it is only possible in
even-dimensional space-times, and the resulting conformal field
(Generalized Maxwell Field, GMF) is represented by
a $(D/2)$-form, where $D=2n$ is the space-time dimension:
\beq                                                        \label{dU} 
	F = dU,\qquad {\rm or} \qquad
	F_{A_1\ldots A_n} = n! \d_{[A_1} U_{A_2\ldots A_n]}
\eeq
where $U$ is a potential $(n-1)$-form and square brackets denote alternation.
The field $F$ is invariant with respect to the gauge transformation
\beq
	U \ \mapsto \ U + dZ                                   \label{dZ} 
\eeq
where $Z$ is an arbitrary $(n-2)$-form; in 4 dimensions this is the
conventional gradient transformation of the Maxwell field.

This combination of gravity and the GMF in
higher dimensions can potentially form the bosonic sector (or part of the
bosonic sector) of some supersymmetric theory. For example, in eight
dimensions the metric tensor --- GMF system has the same number of degrees
of freedom as the Rarita-Schwinger field, and the Lagrangian has a form very
close to the one stemming from truncation of the Lagrangian of $N=1$, $D=11$
supergravity. Fields represented by different $m$-forms like the GMF are
also considered in modern unified models like M-theory --- see
Ref.\,\cite{im96} in this number of the journal.

In the previous papers some cosmological manifestations of the field
(\ref{dU}) were investigated \cite{5,6,7}. One of the main
features of the obtained models is the possible existence of an
initial contracting phase; in some cases, the solutions are
singularity-free.  These results were, of course, obtained with just the
4-dimensional scalar sector of the GMF.

In this paper we analyze $D$-dimensional \sph solutions for the
Einstein-GMF system. We find analytical solutions for many field
configurations, but not for the most general case. In general, we consider
electric, magnetic and scalar charges associated with the GMF. The internal
spaces are assumed to be Ricci-flat.

The great majority of the solutions exhibit naked singularities, but
there are also some classes of \bh\ solutions.
We find the corresponding Hawking temperature and discuss its charge/mass
and dimension dependence. A feature of interest is its possible infinite
growth in the extreme \bh\ limit.

The presence of naked singularities is a common feature of \sph
configurations in multidimensional theories, both in pure gravity and
gravity coupled to gauge and dilaton fields \cite{8,8a,9},
quite similarly to the case of minimal coupling between gravity and scalar
fields in four dimensions. In the latter case the absence of event horizons
is known to be a manifestation of the
so-called no-hair theorems, claiming, in particular, that a \bh\ cannot have
a nonzero scalar charge \cite{mtw}. Likewise,
in pure multidimensional gravity \sph black hole solutions are
only possible when the internal spaces are ``frozen", i.e., the
extra-dimension scale factors, which behave as 4-dimensional scalars,
are constant \cite{br-iv}. No \bh\ solutions exist with a scalar-type
component of a multidimensional Maxwell field \cite{8a,9}. In dilaton
gravity, where special \bh\ solutions exist in the presence of a nontrivial
scalar (dilaton) field, the extra dimensions in the ``string metric" of
these solutions are again ``frozen" (see for instance \cite{9}).
Accordingly, the \bh\ Hawking temperatures in both Einstein an
dilaton gravity does not depend on space-time dimension.
We shall see that such a dependence exists in the case under consideration
in this paper.

A natural next step (going, however, beyond the scope of this
paper) is to study multidimensional configurations with two conformally
invariant fields --- the GMF and the conformal scalar field.
There is a well-known conformal mapping that reduces theories with
nonminimally coupled scalar fields to those with a minimally coupled one
(denoted \phim) --- see \cite{wag} for $D=4$ and, for instance, \cite{bkm}
for $D \geq 4$. Keeping in mind this mapping, we seek all solutions in
the presence of \phim. This addition does not complicate the
solution process and, moreover, the influence of \phim\
upon the properties of the system is also of certain interest.

The paper is organized as follows. In Section 2 we introduce the GMF and
classify the possible GMF configurations compatible with
spherical symmetry. In Sections 3, 4 and 5 we study
the Einstein-GMF-\phim\ system, for each GMF configuration
separately. The Hawking temperature for the
black hole solutions is determined and discussed in Section 6.
Section 7 contains a brief discussion.

Throughout the paper capital Latin indices range over all $D$ coordinates,
Greek ones take the values 0, 1, 2, 3 and small Latin ones refer to extra
dimensions ($i$ is the number of a factor space). The units where
$c=\hbar=1$ are used.

\section{Generalized Maxwell field and spherical symmetry}
We consider general relativity in a Riemannian space-time $V^D$ ($D=2n$,
$n\geq 2$)
in the presence of two minimally coupled massless fields, with the action
\bearr                                                                  
 S = \int  d^D x\sqrt{g}\bigr[R +(-1)^{n-1}F^2 +
          \phi^{,A}\phi_{,A}\bigl],         \nnn              \label{a}
     \cm     F^2 \equiv F^{A_1...A_n}F_{A_1...A_n},
\ear
where $R={}^DR$ is the scalar curvature corresponding to the $D$-metric
$g_{AB}$, $g=|\det g_{AB}|$, $\phi={}$\phim, and $F$ is the GMF.

The corresponding field equations are
\bear
	G^B_A \equiv R^B_A - \half \delta^B_A R \eql -T^B_A,      \label{eqE}\\
                 \nabla_A F^{AA_1\ldots A_n} \eql 0,           \label{eqM}\\
			  \nabla^A\nabla_A \phi       \eql 0.           \label{eqS}
\ear
where the energy-momentum tensor (EMT) $T^B_A$ is a sum of contributions
from \phim\  and $F$. The EMT of \phim\ has its usual form, while that of the
$F$ field is
\beq                            \wider                               
\T{F}^B_A = (-1)^{n-1}\bigl( nF_{AC_2...C_n}F^{BC_2...C_n} -
      \half \delta^B_A F^2\bigr).                              \label{EMT}
\eeq

The $F$-sector of the action is invariant under conformal
mappings of $V^D$ ($g_{AB} \mapsto f(x^A)g_{AB}$) provided the components
$F_{A_1\ldots A_n}$ are unchanged in such transformations
(the potential $U$ from (\ref{dU}) may
experience a gauge transformation (\ref{dZ}) where $Z$ is
some $(n-2)$-form depending on the conformal factor $f$).

Consider $V^D$ with the structure
\beq                                                    	\label{Stru}
V^D = M^4  \times V_1 \times ... \times V_s, \qquad  \dim V_i= N_i,
\eeq
where $s$ is the (so far unspecified) number of Ricci-flat internal spaces;
$s=0$ corresponds to conventional 4-dimensional theory.

The static, \sph metric in $V^D$ may be written in the form
\bearr                           
\ds = \e^{2\gamma(u)}dt^2 -
	\e^{2\alpha(u)}du^2 - \e^{2\beta(u)} d\Omega^2   \nnn
\cm   \cm    	+ \sum_i\e^{2\beta_i(u)}{ds_i}^2,             \label{m}
\ear
where $x^1 = u$ is the radial coordinate and
$d\Omega^2 =d\theta^2 + \sin^2\theta\,d\varphi^2$.

The following nonzero components of $F$ are compatible with
spherical symmetry:
\begin{itemize}
\item $F_{01a_3...a_n}$, \quad an electric field,
\item $F_{23b_3...b_n}$, \quad a magnetic field,
\item $F_{1c_2...c_n}$ , \quad a ``quasiscalar" component, behaving
	as a scalar field in $M^4$,
\end{itemize}
where the indices $a_i$, $b_i$, $c_i$ belong to the internal spaces.

\medskip\noi
{\bf Remark.}\
Several components of each type can exist simultaneously, with different
sets of indices $a$, $b$, $c$; the only restriction is that neither
pair of such sets may contain $n-1$ common indices:
in such cases off-diagonal components of the $F$ field EMT would come into
play, which is incompatible with the Einstein equations for (\ref{m}).

\medskip
We will look for solutions in the following cases:
\begin{itemize}
\item EM\phim: electric and magnetic field components of $F$, plus \phim;
\item S\phim: a quasiscalar component of $F$, plus \phim.
\item EMS\phim: electric, magnetic and quasiscalar field components
	of $F$, plus \phim;
\end{itemize}

Let us choose the harmonic radial coordinate $u$ specified by the condition
\cite{br-acta}
\beq                                             \label{harm}
\alpha = \gamma + 2\beta + \sigma , \qquad \sigma \equiv \sumi N_i\beta_i.
\eeq
Then the nonzero components of the Ricci tensor may be written in the form
\bear
\e^{2\alpha}R^1_1 &=& - \alpha'' + {\alpha'}^2 -
\sum_{i=-1}^s N_i{{\beta_i}'}^2 ,                         \nn
\e^{2\alpha}R^2_2 &=& \e^{2\alpha}{R^3}_3 = \e^{2\alpha - 2\beta}
-\beta'' ,                                               \nn
\e^{-2\alpha}R^{a_i}_{b_i} &=& -\delta^{a_i}_{b_j}{\beta_i}'',\quad
i=-1,1,...,s ,                                                  \label{e3}
\ear
where we have denoted
\beq
\beta_{-1} = \gamma,\ \quad N_{-1} = 1,\ \quad \beta_0 =\beta,\ \quad N_0= 2.
\eeq

The Einstein tensor component $G^1_1$ is
\beq                                                          
\e^{2\alpha}G^1_1 = - \e^{2\alpha - 2\beta} + \Half{\alpha'}^2 -
              \Half \sum_{i=-1}^sN_i{{\beta_i}'}^2            \label{e4}
\eeq
and does not contain second-order derivatives. The corresponding
component of the Einstein equations is an integral to the other components,
similar to the energy integral in cosmology.

The GMF components specified before are easily found from the field equations
(\ref{eqM}):
\bearr                                                   
\begin{array}{lrcrr}
\nhh\mbox {Electric:} & F^{01a_3...a_n} \eql q_\e\e^{-2\alpha},
			       & q_\e = \const ;                      \\
\nhh\mbox {Magnetic:} & F^{23b_3...b_n} \eql q_\m \sin\theta,
			       & q_\m = \const ;                      \\
\nhh\mbox {Quasiscalar:} & F^{1c_2...c_n}\eql  q_\s \e^{-2\alpha},
                      & q_{\rm s} = \const.
		\end{array}		         \label{max}   \nnn
\ear
The EMT (\ref{EMT}), subject to the above restrictions upon the set of
indices, forms a sum of EMTs calculated for each of the components
separately.

It should be noted that $q_\e$ and $q_\m$ are not the physical
electric and magnetic charges in conventional units,
corresponding to the 4-dimensional Maxwell field.
Indeed, the 4-dimensional Einstein-Maxwell Lagrangian (multiplied by
$16\pi G$, $G$ being the Newton constant of gravity) is
\beq
	{}^4\! R - G\F^{\mu\nu}\F_{\mu\nu},                       \label{EMax}
\eeq
where $\F_{\mu\nu}$ is the Maxwell field.
Let us assume that there is a flat space-time asymptotic, where all
extra-dimension scale factors are normalized to unity. Then,
comparing (\ref{EMax}) with the 4-dimensional reduction of (\ref{a})
without \phim, one arrives at the asymptotic identification
\beq
	\F_{\mu\nu} = \sqrt{n!/(2G)} F_{\mu\nu a_3\ldots a_n}   \label{id-F}
\eeq
where the indices $a_j$ correspond to a nonzero component of the GMF.
Accordingly, the charges in conventional units $q_{\rm e\ phys}$ and
$q_{\rm m\ phys}$ are connected with $q_\e$ and $q_\m$ by
\bear                                                     
	q_{\rm e\ phys} \eql \sqrt{n!/(2G)} q_\e,  \nn
	q_{\rm m\ phys} \eql \sqrt{n!/(2G)} q_\m.             \label{id-q}
\ear

The scalar field $\phi$ satisfying the D'Alembert equation (\ref{eqS})
under the above coordinate condition may be written as
\beq
\phi = \phi_0 + \phi_1 u , \quad \phi_0,\phi_1 = \const ,    \label{phi}
\eeq
and its EMT takes the form
\beq
\e^{2\alpha}\T{\phi}^B_A = \phi_1^2\ \diag\left(1, -1,[1]_{D-2}\right).
											   \label{EMT-phi}
\eeq
Here and henceforth the notation $[A]_k$ means that a quantity $A$ is
repeated $k$ times along the diagonal.

Eq.\,(\ref{EMT-phi}) means, in particular,
that if we write the $D$-dimensional Einstein equations in the form
\beq
 R^B_A = - T^B_A + \frac{1}{D-2}\ \delta^B_A\,T^C_C,    \label{eqE'}
\eeq
then a term connected with $\phi$ appears only in the ${1 \choose 1}$
component, which is reasonably replaced by the integral coming from
(\ref{e4}). So the field \phim\  does not affect the equations
for $A$, $B \ne 1$; the presence of \phim\  only modifies a
relation between integration constants appearing from
the equation ${1 \choose 1}$ (the first integral).

\section{Problem EM\phim: integrable cases}             

Given the electric and magnetic type fields, with arbitrary
$n$, the whole set $I$ of internal indices naturally splits into
four subsets, to be labelled by different $i$:
\bear
a \cap b               &\mapsto & i = 1 \ ; \nn
a \setminus b          &\mapsto & i = 2 \ ; \nn
b \setminus a          &\mapsto & i = 3 \ ; \nn
I \setminus (a \cup b) &\mapsto & i = 4 \
\ear
where $a =\{a_k\}$ and $b = \{b_k\}$ are the sets of indices of the nonzero
``electric" and ``magnetic" components (\ref{max}) of the tensor $F$.

It is easy to verify that the corresponding dimensions $N_i$ are
connected by the constraints
\bearr
N_1 + N_2 = N_3 + N_4 = n - 2 ; \nnn
N_1 = N_4 ; \qquad\ N_2 =N_3.
\ear
Accordingly, the EMT of the $F$ field takes the form
\bearr                                                    
\T{F}^B_A = Q_\e^2\,\e^{2x}
     \diag\left(1,1,-1,-1, \pO_{n-2},\mO_{n-2}\right)           \nnn
\nq + Q_\m^2 \e^{2y}\diag\left(1,1,-1,-1, \mO_{\ss N_1}, \pO_{\ss N_2},
                \mO_{\ss N_3},\pO_{\ss N_4}\right)              \nnn
\ear
where
\bear              
Q_\e^2 \eql \half n!\,q_\e^2 , \qquad\ Q_\m^2 = \half n!\,q_\m^2;  \nnv
      x \eql \gamma + N_1\beta_1 + N_2\beta_2 ,                  \nn
      y \eql \gamma + N_2\beta_2 + N_4\beta_4 .
\ear
(one sees that the $Q$'s coincide with the ``physical" charges (\ref{id-q})
up to $\sqrt{G}$).
With these expressions the Einstein equations ($A,B \ne 1$) read:
\bear
\e^{2\alpha - 2\beta} - \beta'' \eql  Q_\e^2\,\e^{2x} +Q_\m^2\,\e^{2y}; \nn
\gamma'' \eql  {Q_\e}^2\e^{2x} + Q_\m^2\,\e^{2y} ;         \nn
{\beta_1}'' \eql  Q_\e^2\e^{2x} - Q_\e^2\e^{2y};       \nn
{\beta_2}'' \eql  \quad \gamma'' ;                               \nn
{\beta_3}'' \eql  - \gamma'' ;                             \nn
{\beta_4}'' \eql  - {\beta_1}'' .                  \label{de6}
\ear

Certain combinations of these equations can be written in terms of
$(\alpha - \beta)$, $x$ and $y$:
\bear                                
(\alpha \al - \al \beta)'' =  \e^{2\alpha - 2\beta} ;         \label{de7}\\
x'' \eql  (n-1)Q_\e^2\e^{2x} + (1 - N_1 + N_2)Q_\m^2\e^{2y}; \label{de8}\\
y'' \eql  (1-N_1+N_2)\e^{2x} + (n-1){Q_\m}^2\e^{2y} .        \label{de9}
\ear
\Eq{de7} immediately gives
\beq                                     
\e^{\beta - \alpha} = s(k,u)
\eeq
	where the function $s(k,u)$ is defined as follows:
\beq                                         
	s(k,u) \equiv \vars{ (1/k)\sinh ku,  \quad & k>0,\\
			      			   u, &        k=0,\\
				      (1/k)\sin ku,  &        k<0.}    \label{Def-s}
\eeq

Here $k = \const$ and another integration constant is
suppressed by a choice of the origin of $u$.

Eqs.\,(\ref{de8}) and (\ref{de9}) decouple and are then easily
solved in the following cases:

\begin{description}
\item[(i)] $N_1=N_4=0$, that is, $a \cap b =\O$. This happens when the
electric and magnetic fields are specified by mutually dual
components of the GMF, as
is the case in conventional 4-dimensional electrodynamics.
Then the functions $x$ and $y$ coincide and the equations
for $\beta_1$ and $\beta_4$ disappear, since the corresponding
subspaces are absent. This defines what we call the EM(dual)-\phim\  system.
\item [(ii)] $N_1 = 1 + N_2$. Since $N_1 + N_2 = n-2$, one can write:
\bearr
n = 2m + 1 , \qquad m = 1,2,... ; \nnn
N_1 = N_4 = m ; \ \ \qquad N_2 = N_3 = m-1 .              \label{ii}
\ear
So this solvable case occurs with the dimensions $D = 2n = 6, 10,...$.
This defines another integrable case, labelled EM(non-dual)-\phim.
\end{description}

For $D = 6$ one has only two variants,
either (i), with $F_{015} \ne 0$ and $F_{236} \ne 0$, or (ii) with
$F_{016} \neq 0$ and $F_{235} \neq 0$. For $n>3$ ($D>6$) there exist other
variants, when our equations are not so easily (if at all) integrable.

\subsection{Solutions EM(dual)-\phim}

In the previously described case (i) we have
\bearr
x''\equiv \gamma'' + (n-2){\beta_2}'' = Q^2\e^{2x} , \nnn
    \cm Q^2 = (n-1)({Q_\e}^2 + {Q_\m}^2),
\ear
whence
\beq
\e^{-x} = Q s(h_1,u+u_1)                                \label{x}
\eeq
where $h$ and $u_1$ are constants and the function $s$ is defined
in (\ref{Def-s}). It is convenient to normalize all the functions
$\gamma$ and $\beta_i$ by the conditions
\beq
\gamma(0) = \beta_i(0) = 0 ,
\eeq
i.e., at the flat-space asymptotic all the exponential functions in the
metric tend to unity (the real sizes of the internal spaces are thus hidden
in ${ds_i}^2$).  Then, in particular,
\beq
	Qs(h_1,u_1) = 1.
\eeq

Further simple integration finally gives
\bearr
\ds = \e^{2\gamma}dt^2 - \frac{\e^{2\beta_0}}{s^2(k,u)}
\biggr[\frac{du^2}{s^2(k,u)} + d\Omega^2\biggl] \nnn
\cm +
\e^{2\beta_2}{ds_2}^2 + \e^{2\beta_3}{ds_3}^2 ,              \label{m1}
\ear
with (\ref{x}) and
\bear
(n-1)\gamma\ \eql  \quad x - (n-2)h_2u ,  \nn
(n-1)\beta_2 \eql  \quad x + h_2u ,      \nn
(n-1)\beta_3 \eql  - x + h_3u ,    \nn
(n-1)\beta_0 \eql  - x - (n-2)h_3u \label{f4}
\ear
where $h_i = \const$; the $F$ field components are
\bear
F_{01a_2...a_n} \eql  (-1)^{n-1}q_\e\e^{2x} , \nn 
F_{23b_1...b_n} \eql  q_\m \sin\theta .\label{F2}
\ear

A substitution of (\ref{m1}), (\ref{f4}) to the constraint equation leads to
the following relation between the integration constants:
\bearr                                           
\nhq 2k^2\sign k = \frac{2}{n{-}1}h_1^2\sign h_1 +
\frac{n{-}2}{n{-}1}(h_2^2 + h_3^2) + 2\phi_1^2 ,   \label{Int-1} \nnn
\ear
so that there are six independent parameters in the solution:
$q_1$, $q_2$, $h_1$, $h_2$, $h_3$ and ${\phi_1}^2$.

The solution generalizes the \RN one and reduces to the latter in the case
$n = 2$ (when the internal spaces disappear), $\phi_1=0$.

An analysis of the solution for $n \geq 3$ reveals naked singularities
in most cases, but under some special conditions, in the absence of
a $\phi$ field, one finds a \bh. Indeed, if
\bearr
\phi_1 = 0 , \qquad k = h_1 = h_2 = - h_3,   \nnn
\cm \qquad u_1 > 0 , \qquad k > 0
\ear
\Eq{Int-1} is satisfied, the functions
$\beta $, $\beta_1$, $\beta_2$ remain finite as $u \to \infty$,
while $\e^\gamma \to 0$, and the light travel time $\int \e^{\alpha -
\gamma}du$, taken from any finite $u$ to $u =\infty$, diverges.

The corresponding black-hole metric looks simpler after the
transformation $u \mapsto R$
\beq     \wider
\e^{-2ku} = 1 - 2k/R,                     \label{tbh1}
\eeq
so that $u \to \infty$ corresponds to $R \to 2k$. One obtains
\bearr                              
\nq  \ds = \frac{1{-}2k/R}{(1{+}p/R)^\xi}dt^2 -
(1{+}p/R)^\xi\Bigl(\frac{dR^2}{1{-}2k/R}-R^2 d\Omega^2\Bigr)  \nnn
\qquad +
\frac{{ds_2}^2}{(1{+}p/R)^\xi} + (1{+}p/R)^\xi{ds_3}^2   \label{mbh1}
\ear
where
\beq
\xi = 2/(n-1) , \qquad p = \sqrt{k^2 + Q^2} - k . \label{cbh1}
\eeq
In the absence of charges ($Q = 0$, $p = 0$) the solution
becomes Schwarzschild's, with $GM = k$ ($G$ is the gravitational constant
and $M$ is the mass), with trivial extra dimensions.

In the general case the gravitating mass of the configuration
is obtained from a comparison with the Schwarzschild metric, so that
\beq
GM = k + p/(n-1) .                                 \label{M1}
\eeq
The charge combination $Q$ is restricted by
\beq
Q^2 < (n-1)^2G^2 M^2 .                                   \label{QM1}
\eeq

In the extreme case $|Q| = GM(n-1)$, $k = 0$, the horizon
$R = 2k$ disappears and, as is easily seen from (\ref{mbh1},
\ref{cbh1}), a naked singularity occurs at $R = 0$, the center of symmetry.
An exception is the ``old" case $n=2$, $D=4$, when (\ref{mbh1}) acquires the
familiar \RN form after the substitution $R + p = r$.

\subsection{Solution EM(non-dual)-\phim}
Now consider the case (ii). It is slightly more complex
since there are four internal spaces for $n > 3$.
With (\ref{ii}), \eqs (\ref{de8}) and (\ref{de9}) yield
\bear
\e^{-x} \eql  \sqrt{2m Q_\e^2}\ s(h_1, u + u_1) , \nn
\e^{-y} \eql  \sqrt{2m Q_\m^2}\ s(h_2, u + u_2) ,
\ear
and, after simple integration of the remaining equations, the
solution takes the form
\bearr                                
	\ds = \e^{2\gamma}dt^2 - \frac{\e^{2\beta_0}}{s^2(k,u)}
	\biggr[\frac{du^2}{s^2(k,u)} + d\Omega^2\biggl]     \nnn
\cm \cm + \sum_{i=1}^4 \e^{2\beta_i(u)}{ds_i}^2 ,        \label{m2}
\ear
	with
\bear
\gamma  \eql  \ (x + y)/(2m) - c_0 u , \nn
\beta_1 \eql  \ (x - y)/(2m) - c_1 u , \nn
\beta_2 \eql  \ (x + y)/(2m) - c_2 u , \nn
\beta_3 \eql  - (x + y)/(2m) - c_3 u , \nn
\beta_4 \eql  \ (y - x)/(2m) - c_1 u , \nn
\beta_0 \eql  - (x + y)/(2m) + mc_1 u + (n - 1)c_3 u ,    \label{s6}
\ear
where the integration constants $k$, $h_i$ and $c_i$ are connected
by the relations
\bearr
     c_0 + mc_1 + (m - 1) = 0 , \label{c1}\yyy
k^2  \sign k = \frac{1}{m}({k_1}^2\sign h_1 + h_2^2\sign h_2) +c^2_0 \nnn
\cm + [mc_1 - (m - 1)c_3]^2 + 2m c_1^2    \nnn
\cm + (m - 1)(c_2^2 + c_3^2) + 2\phi_1^2 .                 \label{c2}
\ear
The $F$ field components retain the same form (\ref{F2}). Thus there are
eight independent integration constants:
($q_\e$, $q_\m$, $h_1$, $h_2$, $c_1$, $c_2$, $c_3$, $\phi_1$).

Like that of Subsec.\,3.1, this solution possesses naked singularities
in most cases, but there is a 3-parameter set of black holes specified by
the conditions
\bearr
	\phi_1 = 0 , \qquad c_1 = 0 ,\nnn
	h_1 = h_2 = k , \qquad -c_2 = c_3 = k/m .
\ear
Indeed, after the same transformation (\ref{tbh1}) the metric looks as
follows:
\bearr     
\ds = \frac{1 - 2k/R}{(P_\e P_\m)^{1/m}}dt^2         \nnn
\nhq -(P_\e P_\m)^{1/m}\biggr(\frac{dR^2}{1 - 2k/R}+R^2 d\Omega^2\biggl)
		+ \Bigl(\frac{P_\m}{P_\e}\Bigr)^{1/m}ds_1^2   \nnn
\nhq + (P_\e P_\m)^{-1/m}ds_2^2  + (P_\e P_\m)^{1/m}ds_3^2
    + \Bigl(\frac{P_\e}{P_\m}\Bigr)^{{1/m}}ds_4^2     \label{mbh2}  \nnn
\ear
with
\bear                                              \label{p2}  
 P_{\e,\m} \eql 1 + p_{\e,\m}/{R} ; \nn
 p_{\e,\m} \eql \sqrt{2mQ_{\e,\m}^2 + k^2} - k.
\ear
Like (\ref{mbh1}), the metric (\ref{mbh2}) has the property
\beq                                                         \label{sig2}
\sigma \equiv \sum_i N_i\beta_i \equiv 0 ,
\eeq
so that the volume element of the extra dimensions is $R$-independent.

The gravitating mass is calculated for (\ref{mbh2}) similarly to
(\ref{mbh1}):
\beq
GM = k + \frac{1}{2m}(p_\e + p_\m) .                         \label{M2}
\eeq
The charge parameters $Q_\e$ and $Q_\m$ are constrained by
\beq                                                         \label{QM2}
Q_\e^2 + Q_\m^2 < 2mG^2 M^2 .\label{cbh2}
\eeq
Again, in the extreme case $k = 0$, when we have an equality in (\ref{cbh2}),
the metric has a singular center $R = 0$.

Unlike (\ref{mbh1}), (\ref{mbh2}) contains three instead of two
parameters since the electric and magnetic charges enter into it separately
(although in a symmetric manner).  Despite the close similarity between
these metrics, (\ref{m2}) has no \RN special case, since the total
space-time dimension is $D = 2(2m+1)$, $m \geq 1$. For $m = 1$, $n = 3$, the
subspaces with ${ds_2}^2$ and ${ds_3}^2$ are absent: by (\ref{ii}) they have
zero dimension.

\section{Solutions with a scalar-type component}

Let us now try to find solutions containing the scalar component of the
$F$ field. We will also assume that there are mutually dual electric
and magnetic fields, so that their sets of indices are completely
different, $a \cap b = \O$. We ascribe the
folowing labels $i$ to the internal factor spaces
(where $a = \{a_j\}$, etc., see (\ref{max}):
\bear
a \cap c      &\mapsto& i = 1 ; \nn
a \setminus c &\mapsto& i = 2 ; \nn
b \setminus c &\mapsto& i = 3 ; \nn
b \cap c      &\mapsto& i = 4 .
\ear
The EMT of the $F$ field takes the form
\bearr
\nq \e^{2\alpha}\T{F}^B_A = Q_{\rm em}^2 \diag(1,1,-1,-1,[1]_{n-2},
                                                 \mO_{n-2})\e^{2x}  \nnn
 + {Q_s}^2\diag(1,-1,1,1,\mO_{N_1},[1]_{N_2+N_3},\mO_{N_4})\e^{2y} \nnn
\ear
where
\bear
Q_{\rm em} \eql  \half n!\ ({q_\e}^2 + {q_\m}^2) , \qquad
Q_{\rm s}  =     \half n!\ {q_s}^2 ,       \nn
x \eql  \gamma + N_1\beta_1 + N_2\beta_2 , \nn
y \eql  N_1\beta_1 + N_4\beta_4 .
\ear
Denoting $N_1 = m \geq 1$, we obtain the following $N_i $:
\bear                                                     \label{60}
N_1 \eql m , \qquad N_2 = n - m - 2 , \nn
N_3 \eql m - 1,   \cm		\quad N_4 = n - m - 1 .
\ear
The field equations similar to (\ref{de6}) are
\bear
\e^{2\alpha - 2\beta} - \beta''
		\eql  {Q_{\rm em}}^2\e^{2x} - {Q_\s}^2\e^{2y},\nn
\gamma''  \eql  {Q_{\rm em}}^2\e^{2x} + {Q_\s}^2\e^{2y} , \nn
\beta_1'' \eql  {Q_{\rm em}}^2\e^{2x} - {Q_\s}^2\e^{2y} ,\nn
\beta_2'' \eql  \ \ \gamma'',    \nn
\beta_3'' \eql  - \beta_1'',     \nn
\beta_4'' \eql  - \gamma''.      \label{e6'}
\ear
As in Subsection 3.1, we obtain (\ref{de7}), while the equations
for $x$ and $y$ are
\bear
x'' \eql (n-1)Q_{\rm em}^2\e^{2x} + (n - 2m - 1){Q_\s}^2\e^{2y} ,\nn
y'' \eql -(n-2m-1)Q_{\rm em}^2\e^{2x} - (n-1)Q_\s^2\e^{2y}. \label{xy1}
\ear

The situation is somewhat similar to that in
equations (\ref{de8},\ref{de9}), with the difference that we cannot put
$x = y$: the set of equations (\ref{e6'}) then becomes
inconsistent. So the only case when \eqs(\ref{xy1}) decouple is
\beq
n = 2m + 1 , \quad m = 1,2,...
\eeq
and we again deal only with $D = 2n = 6, 10, 14,...$, so that, by
(\ref{60}),
\[
  N_1=N_4=m, \qquad N_2=N_3=m-1.
\]
It is easy to verify that the case of coinciding indices indicated in the
Remark in \sect 2 ($a\cap c=a$) occurs here only for $m=1,\ n=3$. In this
case our solution cannot contain an electric component $F_{01a}$ and
incorporates only magnetic and scalar charges.

After integration we obtain
\bear
\e^{-x} \eql  \frac{1}{h_1}\sqrt{n - 1}{Q_{\rm em}}^2s(h_1, u + u_1) ,\nn
\e^{-y} \eql  \frac{1}{h_2}\sqrt{n - 1}{Q_\s}^2\cosh[h_2(u + u_2)] \quad,
\ear
where, as before, $h_1$, $h_2$, $u_1$, $u_2$ are constants constrained
by the conditions $x(0) = y(0) = 0$.

Further integration is quite simple and results in a metric having the
form (\ref{m2}), but with the following exponents instead of (\ref{s6}):
\bear                           
\gamma  \eql  \ \ {x - y}/(2m) - c_0 u , \nn
\beta_1 \eql  \ \ {x + y}/(2m) - c_1 u , \nn
\beta_2 \eql  \ \ {x - y}/(2m) - c_2 u , \nn
\beta_3 \eql  -   {x + y}/(2m) - c_3 u , \nn
\beta_4 \eql  \ \ {y - x}/(2m) + c_1 u , \nn
\beta_0 \eql  - {x + y}/(2m) - mc_1u + (m - 1)c_3u ,
\ear
with constraints upon the constants very similar to (\ref{c1}) and (\ref{c2}):
\bearr
c_0 + mc_1 + (m - 1)c_2 = 0 , \yyy
2k^2\sign k = \frac{1}{m}(h_1^2 \sign h_1 + h_2^2) + c_0^2 + 2m c_1^2  \nnn
\nhq + [mc_1 - (m -1)c_3]^2 + (m - 1)(c_2^2 + c_3^2) + 2\phi_1^2 .
\ear

Unlike the solutions \ EM(dual)-\phim\  and \ EM(non-dual)-\phim, this one,
labelled \ EM(dual)-S\phim, does not contain a black-hole case.
Indeed, as is easily verified, all solutions with the coordinate $u$
specified in a finite range $0 < u < \umx < \infty$
(this happens when $h_1 < 0$ and/or $u_1 < 0$) have naked singularities.
As for $u \to \infty$, the place for a horizon in the previous solutions,
the requirement that $\beta_1,...,\beta_4$ tend to finite limits
cannot be fulfilled since now $\beta_1(u) + \beta_4(u) = y/m$,
while $y \to - \infty$ as $u \rightarrow \infty$.

In the case $m = 1$, $n = 3$, the functions $\beta_2$ and $\beta_3$
disappear along with the corresponding factor spaces, and the solution
is considerably simplified.

\section{Solutions S\phim}

Equations (\ref{e6'}) are easily solved for any $n$ if the
electric and magnetic components of the $F$ field are absent.
In this case, the factor spaces ($1$ and $4$) and ($2$ and $3$)
unify and the resulting solution, obtained just as the previous
ones, has the form,
\bearr
\ds= \e^{2\gamma}dt^2 - \frac{\e^{2\beta_0}}{s^2(k,u)}\biggr[
		\frac{du^2}{s^2(k,u)} + d\Omega^2\biggl] \nnn
\cm + \e^{2\beta_1}{ds_1}^2 + \e^{2\beta_2}{ds_2}^2
\ear
where
\bear
\e^{-(n-1)\beta_1} \eql \frac{1}{h}\sqrt{(n-1)Q_\s^2}\cosh[h(u+u_1)],\nnn
         \qquad \beta_1(0)=0, \qquad k > 0 ,\nn
\gamma  \eql  - \beta_1 - c_0u,             \nn
\beta_2 \eql  - \beta_1 - c_2u,             \nn
\beta_0 \eql  - \beta_1 + [c_0 + (n - 3)c_2]u .
\ear
The constants are related by
\bearr
2k^2\sign k = \frac{2}{n-1} h^2 + c_0^2 + [c_0 + (n-3)c_2]^2   \nnn
    \cm        + (n - 3){c_2}^2 + 2{\phi_1}^2.
\ear
There is no black-hole case in this solution as well since in all cases
$\beta_1 \to - \infty$ as $u \to \infty$.

\section{The Hawking temperature for black-hole solutions}

Event horizons are known to induce quantum vacuum instability,
leading to creation of particle-antiparticle pairs \cite{11,12},
also interpreted as \bh\ evaporation. The latter is observable from
infinity as blackbody radiation, whose temperature,
called the Hawking temperature, is thus one of the key parameters of a \bh.
In non-black-hole cases the notion of a temperature is apparently
unapplicable.

According to \cite{12}, the Hawking temperature of a \sph \bh\ can be found
in the form
\bear                                                \label{TBH} 
k_{rm B}T \eql \hbar\ae/2\pi, \nn
      \ae \eql\frac{(\sqrt{g_{00}})'}{\sqrt{-g_{11}}}\Bigg|_{\rm horizon}
				= \e^{\gamma-\alpha}\gamma'\Big|_{\rm horizon},
\ear
where $k_{\rm B}$ is the Boltzmann constant and the notations $\alpha$,
$\gamma$ and ``prime" correspond to \Eq{m}.

It should be noted that the expression (\ref{TBH}) is not only
invariant under reparametrization of the radial coordinate (as is
necessary for any quantity having a direct physical meaning),
but also {\bf conformal gauge independent}, or, in other words, invariant
under conformal mappings of the 4-dimensional metric provided the conformal
factor is regular on the horizon. Indeed, due to the above
reparametrization invariance, we may safely assume that the horizon takes
place at a finite value of the radial coordinate. Then, from
$\e^{\gamma}\to 0$ it follows  $\gamma'\to \infty$ on the horizon, and in
the expression (\ref{TBH}) any finite contribution to $\gamma'$ may be
ignored; but, on the other hand, a regular conformal factor results in
just a finite contribution to $\gamma'$ and does not affect the expression
$\e^{\alpha-\gamma}$ at all.

Thus a \bh\ has the same temperature for observers using different sets of
instruments (such that they see the space-time in different
conformal gauges). The same is true for \bh\ electric and magnetic charges
(if any) but not for the mass, which is determined by the
4-dimensional metric at the asymptotic and is thus sensitive to conformal
factors. In particular, the mass dependence of the temperature is conformal
gauge dependent.

We have obtained here two \bh\ solutions (\ref{mbh1}), (\ref{mbh2}). Let us
find their temperatures, comparing them with some other known solutions
describing multidimensional \bh s. We begin with referring to known results.

\subsection* {Conventional Einstein-Maxwell fields in $D$ dimensions}

The metric of an electrically charged \bh\ may be written
in the form \cite{br-ann,13}
\bearr                                            \label{mbh3}  
\ds = \frac{1-2k/R}{(1+p/R)^2}dt^2              \nnn
  \nq	 - (1+p/R)^{2/N}\biggl[\frac{dR^2}{1-2k/R}
		+ R^2 d\Omega^2 -\sumi ds_i^2 \biggr]
\ear
where the integration constant $k \geq 0$ has the same meaning as in the
present paper,
\bear
N=D-3,\qquad p \eql \sqrt{k^2 + \qo^2} - k, \nn
         \qo^2 \eql 2N q^2/(N+1)
\earn
and $q$ is the electric charge. According to (\ref{TBH}),
\beq            
	  \ae = \frac{1}{4k}\biggl(1 + \frac{p}{2k}\biggr)^{-(N+1)/N}.
\eeq
On the other hand, the black-hole mass is (up to the factor $G$)
$ M = p+k = \sqrt{k^2 + \qo^2} $,
so that $\ae$ can be expressed in terms of the mass and the charge:
\bearr                
 \ae = \half (2\sqrt{M^2-\qo^2})^{1/N}(M+\sqrt{M^2-\qo^2})^{-(N+1)/N}.\nnn
\ear
The minimal mass for a given charge $q$ corresponds to $k=0$ and zero
temperature (the extreme case). For $N>1$ ($D>4$) the metric possesses a
naked singularity instead of a horizon. In other respects the above formulas
are direct generalizations of those well-known for the \RN\ case.

\subsection* {Dyon \bh s in multidimensional dilaton gravity}

In the case of dilaton gravity, known as a field limit of string theory,
the most general \sph \bh\ solution contains both electric ($q_\e$) and
magnetic ($q_\m$) charges; the string metric (i.e., the metric in the string
conformal gauge, fundamental for the underlying theory) reads \cite{9}:
\bearr                                                           \label{mbh4}
    \ds = \frac{(r-r_+)(r-r_\m)}{(r+r_\e-r_\m)^2} dt^2   \nnn
\qquad -\frac{r^2 dr^2}{(r-r_+)(r-r_\m)} - r^2 d\Omega^2 + \sumi ds_i^2
\ear
with the notations
\[
    r_+ = 2k+r_\m \geq r_\m,        \qquad
    r_{\e,\m} = \sqrt{k^2 + 2q_{\e,\m}^2} -k.
\]
    The event horizon takes place at $r=r_+$, except for the
    purely magnetic extreme case $k=r_\e =0$, when the metric is regular
    and $g_{00}=\const$.

    The ``temperature factor" $\ae$ is
\beq                                                           \label{T4}
    \ae = \frac{k}{r_+(2k + r_\e)}
        = \frac{r_+ - r_\m}{r_+(r_+ + r_\e - r_\m)}.
\eeq
    It is dimension-independent and vanishes for $k=0$.

\subsection* {Solutions EM(dual), \Eq{mbh1} and EM(non-dual), \Eq{mbh2}}

The field \phim\ is omitted from the notation since the scalar field
is absent in the \bh\ solutions.

In the dual case the metric and the mass are determined by (\ref{mbh1}) and
(\ref{M1}).  The black-hole ``temperature factor" $\ae$ is
\beq                                                             \label{T1}
	\ae = \frac{1}{4k}\biggl(1 + \frac{p}{2k}\biggr)^{-2/(n-1)}.
\eeq
Recall that here $p = \sqrt{k^2 + Q^2} -k$, $k > 0$ and
$Q^2 = (n-1)(Q^2_\e + Q^2_\m)$, $Q_\e$ and $Q_\m$ being the physical
electric and magnetic charges of the \bh. The well-known result for a \RN
\bh\ \cite{12} is recovered when $n=2$.

In the non-dual case the corresponding relations are (\ref{mbh2}) and
(\ref{M2}). For $\ae$ we obtain:
\beq                                                             \label{T2}
\ae = \frac{1}{4k}\biggl(1 + \frac{p_\e}{2k}\biggr)^{-1/m}
			   \biggl(1 + \frac{p_\m}{2k}\biggr)^{-1/m}.
\eeq
The quantities $p_{\e,\m}$ are expressed in terms of $k>0$ and the charges
in \Eq{p2}. Recall that here the space-time dimension is $D=2(2m+1)$, $m$
being a positive integer, so that this family of solutions does not include
the \RN one.

All the above expressions for $\ae$, except (\ref{T4}), depend
explicitly on the space-time dimension, so that the temperature tends to
a finite limit as $D\to\infty$. In the dilaton case (\ref{mbh4}) the extra
dimensions are ``passive" not only in that their scale factors are constant
(in the string gauge), but also in that the $D$-dependence falls out of
the whole metric. In this respect the dilaton case is the most similar to
the uncharged (Schwarzschild) solution --- the zero charge limit of all
the above solutions.

The expressions for $\ae$ exhibit the most interesting distinctions in the
extreme limit $k\to 0$, describing the minimum possible mass for given
charges. Note that this extreme case corresponds to a
\bh\ only in the \RN case ($D=4$) and for the dilaton solution (\ref{mbh4})
with $q_\e \ne 0$.

Thus, for the known solutions (\ref{mbh3}) and (\ref{mbh4}), the temperature
$T$ vanishes in this extreme case. Unlike that, the expression (\ref{T1})
for Solution EM(dual) vanishes as $k\to 0$ only for $n=2$ (the \RN \bh);
for $n=3$ the limiting $T$ is finite;
for greater $n$, $T\to \infty$ as $k\to 0$. The same picture is
observed with the non-dual case: the expression (\ref{T2}) tends to a finite
limit as $k\to 0$ only if $m=1$ and there is just one nonzero charge, either
electric, or magnetic one (this case actually coincides with a special case
of EM(dual), $n=3$); in all other cases of EM(non-dual) we have $T\to\infty$
as $k\to 0$. So in our case the \bh\ evaporation dynamics should be
drastically different from that of \RN or dilaton \bh s.

\section{Discussion}

We have found some exact \sph static solutions for a
theory containing gravity and a generalized Maxwell field in higher
dimensions in integrable cases which we were able to select.

All the solutions found, except for some special cases,
exhibit naked singularities. However, from the standpoint of the well-known
no-hair theorems (claiming, in particular, that in general relativity a
\sph \bh\ cannot have an external minimally coupled scalar field), it is
more surprising that there exist \bh\ subfamilies in our families of
solutions, since the extra-dimension scale factors are
4-dimensional scalars. The latter,
however, are not minimally coupled to matter, and this is apparently a
reason for the appearance of nontrivial scalar fields in multidimensional
\bh\ solutions.  Other examples of such solutions are those known in
multidimensional Einstein and dilaton gravity \cite{br-iv,8a,9,
br-ann,br-vuz91},
where nontrivial scalar (or 4-dimensional scalar) fields exist only in the
presence of a nonzero electric or/and magnetic charge.

The present black-hole solutions are among the simplest ones in the following
sense. There are physically different 4-dimensional formulations of the
same multidimensional theory, corresponding to different conformal gauges
(Einstein gauge, atomic gauge, etc.) --- conformal factors depending on the
volume of the internal space, which, in general, varies from point to point
\cite{br-th,9}. In the present notation, such conformal factors depend on
$\sigma$ (see (\ref{harm})).  One can see that in our black-hole solutions
$\sigma=\const$, so for these solutions all conformal gauges coincide.

For the black-hole solutions found, the Hawking temperture depends both on
electromagnetic charges and masses and on the space-time dimension $D$.
The $D$ dependence is somewhat similar to that in the solution (\ref{mbh3})
for the conventional Einstein-Maxwell system and disappears with
switching-off the gauge fields.

A crucial, and potentially observable, difference between our black-hole
solutions and the ``old" ones, \eqs (\ref{mbh3}) and (\ref{mbh4}), is that
for most of our solutions the black-hole temperature grows infinitely in the
extreme limit (that of minimal mass for given charges) --- for more details
see the previous section.

An issue of importance is the classical stability of static configurations.
From the previous studies of the stability of static vacuum and
electrovacuum solutions in multidimensional Einstein and dilaton gravity it
can be concluded that only black-hole solutions are stable, while those with
naked singularities are catastrophically unstable \cite{br-vuz92,13}.  It
would be of interest to extend this study to the present solutions with the
GMF.

It also makes sense to analyze the weak-field limit of the new solutions to
learn their viability range and to try to predict their observational
manifestations.

As is the case with other \sph solutions, their actual significance
(in particular, the role of naked singularity solutions and their relation
to the cosmic censorship conjecture \cite{14,15}) can be understood only
after a full dynamical study of gravitational collapse.

\subsection*{Acknowledgement}

We thank CNPq, Brazil, for partial financial support. K.B. is also thankful
to the Russian Ministry of Science for partial financial support of this
work and to Departamento de F\'\i sica, UFES, Vit\'oria, for kind
hospitality.

\end{document}